\documentclass[seceq,preprint]{ptptex}
\usepackage{amsmath,amssymb,mathtools}

\newcommand{\llangle}{\langle\!\langle}
\newcommand{\rrangle}{\rangle\!\rangle}
\newcommand{\qq}{\mathcal{S}}
\newcommand{\oo}{\mathcal{O}}

\notypesetlogo                       
\preprintnumber[3cm]{
YITP-16-147}

\title{
Space-time supersymmetry in \\ WZW-like 
open superstring field theory
}

\author{
Hiroshi \textsc{Kunitomo}\footnote{%
E-mail:\  {\tt kunitomo@yukawa.kyoto-u.ac.jp}}
}

\inst{
Center for Gravitational Physics, Yukawa Institute for Theoretical Physics,\\ 
Kyoto University, Kyoto 606-8502, Japan
}

\abst{
We investigate space-time supersymmetry in the WZW-like open superstring
field theory, whose complete action was recently constructed.
Starting from a natural space-time supersymmetry transformation at the linearized level,
we construct a nonlinear transformation so as to keep the complete action invariant. 
Then we show that the transformation satisfies the supersymmetry algebra up to 
an extra transformation, unphysical on the asymptotic string fields.
This guarantees that the constructed transformation in fact acts 
as space-time supersymmetry on the physical S-matrix.
}

\begin{document}

\maketitle

\section{Introduction}

Construction of a complete action including both the Neveu-Schwarz (NS) sector representing
space-time bosons and the Ramond sector representing space-time fermions are a long-standing 
problem in superstring field theory.
While the action for the NS sector was constructed
based on two different formulations, the WZW-like formulation\cite{Berkovits:1995ab} 
and the homotopy-algebra-based formulation,\cite{Erler:2013xta} 
it had been difficult to 
incorporate the Ramond sector in a Lorentz-covariant way. 
%
%
Only recently, however, a complete action has been constructed for the WZW-like
formulation\cite{Kunitomo:2015usa}, and soon afterwards for
the homotopy-algebra-based formulation.\cite{Erler:2016ybs}
Interestingly enough, in these complete actions,
the string field in each sector appears quite asymmetrically. 
In the WZW-like formulation, for example,
the string field $\Phi$ in the NS sector is in the large Hilbert space,
characterizing the WZW-like formulation, but the string field $\Psi$ 
in the Ramond sector is in the restricted small Hilbert space
defined using the picture-changing operators.
Then the question is how space-time supersymmetry is realized 
between these two apparently asymmetric sectors.
The purpose of this paper is to answer this question by explicitly
constructing the space-time supersymmetry transformation in the WZW-like formulation.\footnote{
Space-time supersymmetry in the homotopy-algebra-based formulation has recently been studied
by Erler.\cite{Erler:2016rxg}} 

In the first quantized formulation, space-time supersymmetry is generated 
by the supercharge obtained
by using the covariant fermion emission vertex,\cite{Friedan:1985ge} 
which interchanges each physical state in the NS sector with that in the Ramond 
sector. Therefore, it is natural to expect first that the space-time 
supersymmetry transformation
in superstring field theory is realized as a linear transformation 
using this first-quantized supercharge.\cite{Witten:1986qs} 
We will see, however, that this expectation 
is true only for the free theory, while the action including the interaction terms 
is not invariant under this linear transformation.
We modify it so as to be a symmetry of the complete action, 
and then verify whether the constructed nonlinear transformation satisfies
the supersymmetry algebra. 
%
We find that the supersymmetry algebra holds,
up to the equations of motion and gauge transformation,
only except for a nonlinear transformation.
It is shown, however, that this extra transformation
can also be absorbed into the gauge transformation up to
the equations of motion at the linearized level. 
Under the assumption that the asymptotic condition
holds also for the string field theory, this implies, at least perturbatively,
that the constructed transformation acts as space-time supersymmetry 
on the physical states defined by the asymptotic string fields. 
This guarantees that supersymmetry is realized on the physical S-matrix.\footnote
{We further assume asymptotic completeness in this paper. }


The rest of the paper is organized as follows.
In section 2, we summarize the known results on the complete action 
for the WZW-like open superstring field theory.
In addition, restricting the background to the flat space-time,
we introduce the GSO projection operator, which is essential
to make the physical spectrum supersymmetric.
For later use, some basic ingredients, such as the Maurer-Cartan equations
and the covariant derivatives, are extended to those based
on general derivations of the string product which can be noncommutative.
After this preparation, the space-time supersymmetry transformation is 
constructed in section 3.
Using the first-quantized supercharge, a linear transformation is
first defined so as to be consistent with the restriction in the Ramond sector.
Since this transformation is only a symmetry of the free theory,
we first construct the nonlinear transformation perturbatively 
by requiring it to keep the complete action invariant.
Based on some lower-order results, we suppose the full
nonlinear transformation $\delta_\qq$ in a closed form,
%
and prove that it is actually a symmetry of the action.
In section 4, the commutator of two transformations
is calculated explicitly. 
We show that it provides the space-time translation $\delta_p$,
up to the equations of motion and gauge transformation,
except for a nonlinear transformation $\delta_{\tilde{p}}$
that can be absorbed into the gauge transformation only at the linearized level. 
Thus the supersymmetry algebra holds only on the physical states, 
and hence the physical S-matrix, defined by the asymptotic string fields
under appropriate assumptions on asymptotic properties of the string fields.
Although this extra symmetry is unphysical in this sense, 
it is nontrivial in the total Hilbert space including
unphysical degrees of freedom. It produces further
unphysical symmetries by taking commutators 
with supersymmetries or themselves successively.
We have a sequence of unphysical symmetries
corresponding to the first-quantized charges
obtained by taking successive commutators of the supercharge
and the unconventional translation charge with picture number $p=-1$.
Section 5 is devoted to summary and discussion, and
two appendices are added.  In Appendix~A,
we summarize the conventions for the $SO(1,9)$ spinor 
and the Ramond ground states,
which are needed to identify the physical spectrum
although they do not appear in this paper explicitly.
The triviality of the extra transformation
in the Ramond sector, which remains to be shown, 
is given in Appendix~B. Further nonlinear
transformations obtained by taking the commutator
of two unphysical transformations, 
$[\delta_{\tilde{p}_1},\delta_{\tilde{p}_2}]$
are also discussed. 
All the extra symmetries obtained by taking commutators 
with $\delta_\qq$ or $\delta_{\tilde{p}}$ repeatedly
are shown to be unphysical. 

\section{Complete gauge-invariant action}

On the basis of the Ramond-Neveu-Schwarz (RNS) 
formulation of superstring theory,
an open superstring field is a state in the conformal 
field theory (CFT) 
consisting of the matter sector, the reparametrization
ghost sector, and the superconformal ghost sector. 
We assume in this paper that the background space-time
is ten-dimensional Minkowski space, for which
the matter sector is described by string coordinates 
$X^\mu(z)$ and their
partners $\psi^\mu(z)$ $(\mu=0,1,\cdots,9)$. The 
reparametrization ghost sector and 
superconformal ghost sector are described
by a fermion pair $(b(z),c(z))$ and a boson pair 
$(\beta(z),\gamma(z))$,
respectively. The superconformal ghost sector has 
another description
by a fermion pair ($\xi(z)$, $\eta(z)$) and a chiral 
boson 
$\phi(z)$ \cite{Friedan:1985ge}. The two descriptions 
are related through 
the bosonization relation:
\begin{equation}
 \beta(z)\ =\ \partial\xi(z) e^{-\phi(z)}\,,\qquad
 \gamma(z)\ =\ e^{\phi(z)} \eta(z)\,.
\end{equation}
The Hilbert space for the $\beta\gamma$ system is 
called the small Hilbert space
and that for the $\xi\eta\phi$ system is called the large 
Hilbert space.

The theory has two sectors depending on the boundary condition 
on the world-sheet
fermions $\psi^\mu$, $\beta$, and $\gamma$.
The sector in which the world-sheet fermion obeys an 
antiperiodic boundary
condition is known as the Neveu-Schwarz (NS) sector, 
and describes the space-time
bosons. The other sector in which the world-sheet 
fermion obeys
a periodic boundary condition is known as the Ramond 
(R) sector,
and describes the space-time fermions.
We can obtain the space-time supersymmetric theory by
suitably combining two sectors\cite{Gliozzi:1976qd}.

\subsection{String fields and constraints}

In the WZW-like open superstring field theory, we use the string field $\Phi$ 
in the large Hilbert space for the NS sector.
It is Grassmann even, and has ghost number 0 and 
picture number 0.
Here we further impose the BRST-invariant GSO 
projection\footnote{
This BRST-invariant GSO projection and that for the Ramond sector 
to be introduced
shortly were first given in Ref.~\citen{Terao:1985rw}.
The operators $G_{NS}$ and $G_R$ are none other than 
world-sheet fermion number operators in the total 
Hilbert space including the ghost sectors.}
\begin{equation}
\Phi\ =\ \frac{1}{2}(1+(-1)^{G_{NS}})\, \Phi\,,
\end{equation}
where $G_{NS}$ is defined by
\begin{align}
 G_{NS}\ =&\ \sum_{r>0}(\psi^\mu_{-r}\psi_{r\mu}-\gamma_{-r}\beta_r+\beta_{-r}\gamma_r) - 1
\nonumber\\
\equiv&\ \sum_{r>0}\psi^\mu_{-r}\psi_{r\mu} + p_\phi\qquad (\textrm{mod}\ 2)\,,
\end{align}
with $p_\phi=-\oint\frac{dz}{2\pi i}\partial\phi(z)$.
This is necessary to remove the tachyon and makes the spectrum 
supersymmetric\cite{Gliozzi:1976qd}.

For the Ramond sector, we use the string field $\Psi$ 
constrained on the restricted
small Hilbert space satisfying the conditions\cite{Kunitomo:2015usa}
\begin{equation}
 \eta\Psi\ =\ 0\,,\qquad
 XY\Psi\ =\ \Psi\,, \label{R constraints}
\end{equation}
where $X$ and $Y$ are the picture-changing operator
and its inverse acting on the states
in the small Hilbert space with picture numbers $-3/2$ 
and $-1/2$, respectively. They are defined by
\begin{equation}
 X\ =\ -\delta(\beta_0)G_0 + \delta'(\beta_0)b_0\,,\qquad Y\ =\ -c_0\delta'(\gamma_0)\,,
\label{PCO}
\end{equation}and satisfy
\begin{equation}
 XYX\ =\ X\,,\qquad YXY\ =\ Y\,, \qquad [Q,\,X]\ =\ 0\,.
 \label{xyx}
\end{equation}
The string field $\Psi$ is Grassmann odd, and has ghost number $1$ and picture number $-1/2$.
The picture-changing operator $X$ is BRST exact in the large Hilbert space,
and can be written using the Heaviside step function as $ X=\{Q,\Theta(\beta_0)\}$. 
Here, instead of $\Theta(\beta_0)$, we introduce
\begin{equation}
\Xi\ =\ \xi_0 + (\Theta(\beta_0)\eta\xi_0 - \xi_0)P_{-3/2}
+ (\xi_0\eta\Theta(\beta_0) - \xi_0)P_{-1/2}\,,
\end{equation}
and anew define 
\begin{equation}
 X\ =\ \{Q,\ \Xi\}\,.
\label{X in Ramond}
\end{equation}
This is identical to the one defined in (\ref{PCO})
when it acts on the states in the small Hilbert space with picture number $-3/2$, 
but can act on the states in the large Hilbert space without the restriction on 
the picture number.\cite{Erler:2016ybs}
The operator $\Xi$ is nilpotent ($\Xi^2=0$) and satisfies $\{\eta, \Xi\}=1$
\cite{Erler:2016ybs}, from which, 
with $\{Q,\eta\}=0$,
we can conclude
\begin{align}
 [\eta, X]\ =&\ [\eta,\{Q,\Xi\}]
\nonumber\\
=&\ -[Q,\{\Xi,\eta\}]-[\Xi,\{\eta,Q\}]\ =\ 0\,.
\end{align}
%
%
%
We impose the BRST-invariant GSO projection as
\begin{equation}
 \Psi\ =\ \frac{1}{2}(1+\hat{\Gamma}_{11}(-1)^{G_R})\,\Psi\,, 
\label{GSO Ramond}
\end{equation}
where $G_R$ is given by
\begin{align}
 G_R\ =&\ \sum_{n>0}(\psi^\mu_{-n}\psi_{n\mu}-\gamma_{-n}\beta_n+\beta_{-n}\gamma_n) 
- \gamma_0\beta_0
\nonumber\\
\equiv&\ \sum_{n>0}\psi^\mu_{-n}\psi_{n\mu} + p_\phi + \frac{1}{2}\qquad (\textrm{mod}\ 2)\,.
\end{align}
The gamma matrix $\hat{\Gamma}_{11}$ is
defined by using the zero-modes of the world-sheet fermion $\psi^\mu(z)$ as
\begin{equation}
\hat{\Gamma}_{11}\ =\ 2^5\,\psi^{0}_0\psi^{1}_0\cdots\psi^{9}_0\,.
\label{gamma11}
\end{equation}
We summarize the convention on how the zero modes $\psi^\mu_0$ act on the Ramond ground states
in Appendix \ref{convention}.\footnote{ 
In the context of string field theory, the GSO projections are also needed 
to make the Grassmann properties of string fields $\Phi$ and $\Psi$ consistent with 
those of the coefficient space-time fields.
}

\subsection{Complete gauge-invariant action}

By use of the string fields introduced in the previous subsection,
the complete action for the WZW-like open superstring field theory is
given by\cite{Kunitomo:2015usa}
\begin{equation}
 S\ =\ -\frac{1}{2}\llangle\Psi, YQ\Psi\rrangle
-\int_0^1 dt \langle A_t(t), QA_\eta(t)+(F(t)\Psi)^2\rangle\,,
\label{complete action}
\end{equation}
and is invariant under the gauge transformations
\begin{subequations}\label{full gauge}
\begin{align}
A_{\delta_g}\ =&\ D_\eta\Omega + Q\Lambda 
+ \{F\Psi,F\Xi\{F\Psi,\Lambda\}\}
- \{F\Psi,F\Xi\lambda\}\,,
\label{gauge tf ns}\\
\delta_g\Psi\ =&\ 
-X\eta F\Xi[F\Psi, D_\eta\Lambda] + Q\lambda + X\eta F\lambda\,,
\label{gauge tf r}
\end{align}
\end{subequations}
where we have introduced the one parameter extension $\Phi(t)$ of $\Phi$\ $(t\in[0,1])$  
satisfying the boundary condition $\Phi(1)=\Phi$ and $\Phi(0)=0$, and defined
\begin{equation}
  A_{\mathcal{O}}(t)\ =\ (\mathcal{O} e^{\Phi(t)})e^{-\Phi(t)}\,, 
\end{equation}
with $\mathcal{O}=\partial_t, \eta,$ or $\delta$, which
are analogs of (components) of the right-invariant one form,
satisfying the Maurer-Cartan-like equation
\begin{equation}
 \mathcal{O}_1A_{\mathcal{O}_2}(t)
 -(-1)^{\mathcal{O}_1\mathcal{O}_2}\mathcal{O}_2A_{\mathcal{O}_1}(t)
-[\![A_{\mathcal{O}_1}(t),\,A_{\mathcal{O}_2}(t)]\!]
=\ 0\,,\label{MC}
\end{equation}
where $[\![A_1,A_2]\!]$
is the graded commutator of the two string field $A_1$ and $A_2$\,:
$[\![A_1,A_2]\!]=A_1A_2-(-1)^{A_1A_2}A_2A_1$\,.
Using $A_\eta(t)$, the covariant derivative $D_\eta(t)$ is defined by 
the operator acting on the string field $A$ as
\begin{equation}
 D_\eta(t) A\ =\ \eta A - [\![A_\eta,\, A]\!]\,,
\end{equation}
which is nilpotent: $(D_\eta(t))^2=0$\,.
Then the linear map $F(t)$ 
on a general string field $\Psi$ in the Ramond sector
is defined by
\begin{align}
F(t)\Psi\ 
=&\ \frac{1}{1+\Xi(D_\eta(t)-\eta)}\,\Psi
\nonumber\\
=&\ \Psi + \Xi[\![A_\eta(t)\,, \Psi]\!] 
+ \Xi[\![A_\eta(t),\Xi[\![A_\eta(t), \Psi]\!] ]\!]+\cdots\,.
\label{def F}
\end{align}
The map $F(t)$ has a property that changes $D_\eta(t)$ into $\eta$\,:
\begin{equation}
D_\eta(t)F(t)\ =\ F(t)\eta\,.
\label{important property} 
\end{equation}
Using $F(t)$, we can define a homotopy operator for $D_\eta(t)$
as $F(t)\Xi$ satisfying\cite{Kunitomo:2015usa}
\begin{equation}
 \{D_\eta(t), F(t)\Xi\}\ =\ 1\,,
\label{homotopy relation}
\end{equation}
which trivializes the $D_\eta$-cohomology as well as the $\eta$-cohomology
in the large Hilbert space.
From the definition (\ref{def F}), 
we can show that the homotopy operator $F\Xi$ is BPZ even
\begin{equation}
 \langle F\Xi \Psi_1, \Psi_2\rangle\ =\ (-1)^{\Psi_1}\langle \Psi_1, F\Xi \Psi_2\rangle\,,
\label{BPZ homotopy R}
\end{equation}
and satisfies
\begin{equation}
 \{Q, F\Xi\}A\ =\
FXF\Xi D_\eta A + FX\eta F\Xi A-F\Xi[QA_\eta, F\Xi A]\,,
\label{Q and FXi}
\end{equation}
for a string field $A$\,.
It is useful to note that we can define the projection operators
\begin{equation}
 \mathcal{P}_R\ =\ D_\eta F\Xi\,,\qquad
 \mathcal{P}_R^{\perp} =\ F\Xi D_\eta\,,
\label{proj ramond}
\end{equation}
onto the Ramond string field annihilated by $D_\eta$ and
its orthogonal complement, respectively.

The BPZ inner product in the small Hilbert space $\llangle\cdot,\cdot\rrangle$
is related to that in the large Hilbert space $\langle\cdot,\cdot\rangle$ as
\begin{align}
 \llangle A\,, B\rrangle\ =&\ \langle\Xi A\,, B\rangle\ =\ (-1)^A\langle A\,, \Xi B\rangle
 \nonumber\\
=&\ \langle\xi_0 A\,, B\rangle\ =\ (-1)^A\langle A\,, \xi_0 B\rangle\,,
\label{small to large}
\end{align}
where $A$ and $B$ are in the small Hilbert space, and also in the Ramond sector 
for the equations in the first line.

Using a general variation of the map $F(t)$ on a string field $A$\,, 
\begin{equation}
 (\delta F(t))A\ 
=\ -F(t)(\delta F^{-1}(t))F(t)A\ 
=\ F\Xi[\![\delta A_\eta(t)\,, F(t)A]\!]\,,
\label{variation F}
\end{equation}
a general variation of the action (\ref{complete action}) can
be calculated as\cite{Kunitomo:2015usa}
\begin{equation}
 \delta S\ =\ - \langle A_\delta, QA_\eta+(F\Psi)^2\rangle
- \llangle\delta\Psi, Y(Q\Psi+X\eta F\Psi)\rrangle\,,
\label{general variation}
\end{equation}
from which we find the equations of motion,
\begin{equation}
 QA_\eta + (F\Psi)^2\ =\ 0\,,\qquad
 Q\Psi + X\eta F\Psi\ =\ 0\,.
\label{equations of motion}
\end{equation}

Before closing this section, we generalize several ingredients
for later use. We can define $A_{\mathcal{O}}(t)$
not only for $\mathcal{O}=\partial_t, \eta,$ or $ \delta$\,,
but also for any other derivations of the string product. 
Although such general $\mathcal{O}$'s are not in general commutative,
we assume that they satisfy a closed algebra with respect to the graded commutator
of derivations,
$\{\mathcal{O}_1,\mathcal{O}_2]\ =\ \mathcal{O}_1\mathcal{O}_2 
-(-1)^{\mathcal{O}_1\mathcal{O}_2}\mathcal{O}_2\mathcal{O}_1$\,.
The generalized $A_{\mathcal{O}}(t)$'s satisfy the equation
\begin{align}
 \mathcal{O}_1A_{\mathcal{O}_2}(t)
 -&(-1)^{\mathcal{O}_1\mathcal{O}_2}\mathcal{O}_2A_{\mathcal{O}_1}(t)
- [\![A_{\mathcal{O}_1}(t)\,, A_{\mathcal{O}_2}(t)]\!]
=\ A_{\{\mathcal{O}_1,\mathcal{O}_2]}(t)\,,\label{gen MC}
\end{align}
which reduces to the Maurer-Cartan-like equation (\ref{MC}) 
when $\{\mathcal{O}_1,\mathcal{O}_2]=0$\,.
Using $A_{\mathcal{O}}(t)$, 
we can define the covariant derivative
$D_{\mathcal{O}}(t)$ on a string field $A$ by
\begin{equation}
 D_{\mathcal{O}}(t) A\ =\ \mathcal{O} A 
- [\![A_{\mathcal{O}}(t)\,, A]\!]\,.
\end{equation}
From (\ref{gen MC}), we can show that 
\begin{equation}
[\![D_{\mathcal{O}_1}(t)\,, D_{\mathcal{O}_2}(t)]\!]\
=\
D_{\{\mathcal{O}_1, \mathcal{O}_2]}(t)\,.
\label{generalized D}
\end{equation}

As an analog of the linear map $F(t)$ in the Ramond sector,
we can also define the linear map $f(t)$ 
on a general string field $\Phi$
in the NS sector by
\begin{align}
 f(t)\Phi\ =&\ 
\frac{1}{1+\xi_0(D_\eta(t)-\eta)}\,\Phi
\nonumber\\
=&\ \Phi + \xi_0 [\![A_\eta(t), \Phi]\!]
+ \xi_0 [\![A_\eta(t),\,\xi_0[\![A_\eta(t), \Phi]\!] ]\!]\cdots\,.
\label{f ns}
\end{align}
A homotopy operator for $D_\eta(t)$ in the NS sector is
given by the BPZ even operator $f(t)\xi_0$\,:
\begin{equation}
 \{D_\eta(t),\, f(t)\xi_0\}\ =\ 1\,,\qquad
 \langle f\xi_0 \Phi_1, \Phi_2\rangle\ =\ 
(-1)^{\Phi_1}\langle \Phi_1, f\xi_0 \Phi_2\rangle\,.
\label{BPZ homotopy NS}
\end{equation}
%
We can define the projection operators 
\begin{equation}
 \mathcal{P}_{NS}\ =\ D_\eta f\xi_0\,,\qquad
 \mathcal{P}_{NS}^\perp\ =\ f\xi_0 D_\eta\,,\qquad
 \label{proj ns}
\end{equation}
onto the NS string
field annihilated by $D_\eta$ and its orthogonal complement,
respectively.

\section{Space-time supersymmetry}

Now let us discuss how space-time supersymmetry is realized
in the WZW-like formulation. 
Starting from a natural linearized transformation exchanging the NS string
field $\Phi$ and the Ramond string field $\Psi$, we construct 
a nonlinear transformation that is a symmetry of the complete action 
(\ref{complete action}). We show that the transformation satisfies
the supersymmetry algebra, up to the equations of motion and gauge transformation,
except for an unphysical symmetry.

\subsection{Space-time supersymmetry transformation}

At the linearized level, a natural space-time supersymmetry transformation
of string fields in the small Hilbert space, $\eta\Phi$ and $\Psi$, is given by
\begin{equation}
\delta^{(0)}_{\qq(\epsilon)} \eta\Phi\ 
=\ \qq(\epsilon)\Psi,\qquad 
\delta^{(0)}_{\qq(\epsilon)} \Psi\ 
=\ X\qq(\epsilon)\eta\Phi\,,
\label{restricted linear}
\end{equation}
where 
\begin{equation}
\qq(\epsilon)\ =\
\epsilon_\alpha q^\alpha\ =\ 
\epsilon_\alpha 
\oint\frac{dz}{2\pi i}S^\alpha(z) e^{-\phi(z)/2}\\,
\label{supercharge}
\end{equation}
is the first-quantized space-time supersymmetry charge with the parameter $\epsilon_\alpha$\,.
The spin operator $S^\alpha(z)$ in the matter sector can be constructed 
from $\psi^\mu(z)$ using the bosonization technique \cite{Friedan:1985ge}.
This $\qq(\epsilon)$ is a (Grassmann-even) derivation of the string product,
and is commutative with $Q$, $\eta$ and $\xi_0$\,:
$[Q,\qq(\epsilon)]=[\eta,\qq(\epsilon)]=[\xi_0, \qq(\epsilon)]=0$\,. 
It satisfies the algebra,
\begin{align}
 [\qq(\epsilon_1), \qq(\epsilon_2)]\ 
=&\
\tilde{p}(v_{12})\,,
\label{1st quantized alg}
\end{align}
with $v_{12}^\mu=(\epsilon_1C\bar{\gamma}^\mu\epsilon_2)/\sqrt{2}$\,,
where $\tilde{p}(v)$ is the operator with picture number $p=-1$  defined by
\begin{equation}
\tilde{p}(v)\ =\ v_\mu\tilde{p}^\mu\ =\ - v_\mu\oint\frac{dz}{2\pi i}\psi^\mu(z) e^{-\phi(z)}\,.
\label{p with -1}
\end{equation}
This is equivalent to the space-time translation operator 
$p(v)=v_\mu\oint\frac{dz}{2\pi i}i\partial X^\mu(z)$ 
(center of mass momentum of the string)
in the sense that, for example,\cite{Witten:1986qs}
\begin{equation}
 (p(v)-X_0\tilde{p}(v))\ =\ \{Q, M(v)\}\,,
\label{p tilde p}
\end{equation}%
with
\begin{equation}
 M(v)\ =\ v^\mu\oint\frac{dz}{2\pi i}(\xi(z)-\xi_0)\psi_\mu(z)e^{-\phi(z)}\,.
\label{kernel M}
\end{equation}
%
Note that $M(v)$ does not include $\xi_0$, and so is in the small Hilbert space\,:
$\{\eta, M(v)\}=0$\,.
The algebra (\ref{1st quantized alg}) and the Jacobi identity imply
that $[Q, \tilde{p}(v)]=[\eta, \tilde{p}(v)]=[\xi_0, \tilde{p}(v)]=0$\,.

We frequently omit specifying the parameters explicitly and denote, for example,
$\qq(\epsilon_1)$ by $\qq_1$\,.
Since $\eta\Phi$ and $\Psi$ are in the small Hilbert space containing the 
physical spectrum,
(\ref{restricted linear}) is the transformation law given in Ref.~\citen{Witten:1986qs} 
except that the local picture-changing operator 
at the midpoint is replaced by the $X$ in (\ref{PCO})
so that the transformation is closed in the restricted space. 
As a transformation of $\Phi$ in the large Hilbert space, 
we adopt here that
\begin{equation}
 \delta^{(0)}_{\qq(\epsilon)}\Phi\ 
 =\ \qq(\epsilon)\Xi\Psi\,.\label{linear tf phi}
\end{equation}
This is consistent with (\ref{restricted linear}) but is not unique.
A different choice, however, can be obtained 
by combining (\ref{linear tf phi}) and an $\Omega$-gauge transformation, 
for example,
\begin{align}
\tilde{\delta}_{\qq(\epsilon)}^{(0)}\Phi\ =&\ \xi_0\qq(\epsilon)\Psi
\nonumber\\
=&\ \delta_{\qq(\epsilon)}^{(0)}\Phi - \eta(\xi_0\qq(\epsilon)\Xi\Psi)\,.
\end{align}
Using the fact that $\qq$ is BPZ odd,
\begin{equation}
\langle \qq A, B\rangle\ =\ -\langle A, \qq B\rangle\,,
\label{BPZ S}
\end{equation}
it is easy to see that 
the quadratic terms of the action (\ref{complete action}),
\begin{equation}
 S^{(0)}\ =\ - \frac{1}{2} \langle\Phi, Q\eta\Phi\rangle
- \frac{1}{2} \llangle\Psi, YQ\Psi\rrangle\,,
\label{kinetic}
\end{equation}
are invariant under the transformation
\begin{equation}
 \delta_\qq^{(0)}\Phi\ =\ \qq\Xi\Psi\,,\qquad
 \delta_\qq^{(0)}\Psi\ =\ X\qq\eta\Phi\,.
\label{linearized tf}
\end{equation}
However, the action at the next order,
\begin{equation}
 S^{(1)}\ =\ -\frac{1}{6}\langle\Phi, Q[\Phi, \eta\Phi]\rangle
- \langle\Phi, \Psi^2\rangle\,,
\end{equation}
is not invariant under $\delta^{(0)}_\qq$ but is transformed as
\begin{align}
 \delta^{(0)}_\qq S^{(1)}\ =\ 
\langle\left(\frac{1}{2}[\Phi, \qq\Xi\Psi]
-\qq\Xi[\Phi, \Psi]
+\{\Psi, \Xi\qq\Phi\}\right),Q\eta\Phi\rangle
\nonumber\\
+\, \llangle\left(-\frac{1}{2}X\eta[\Phi,\qq\Phi]
+X\eta[\Phi,\Xi\qq\eta\Phi]\right), YQ\Psi\rrangle\,.
\label{var one}
\end{align}
We have thus to modify the transformation by adding
\begin{align}
 \delta^{(1)}_\qq \Phi\ =&\ \frac{1}{2}[\Phi, \qq\Xi\Psi]
 -\qq\Xi[\Phi,\Psi]+\{\Psi,\Xi \qq\Phi\},\\
 \delta^{(1)}_\qq \Psi\ =&\ -\frac{1}{2}X\eta[\Phi,\qq\Phi]
 +X\eta[\Phi,\Xi \qq\eta\Phi]\,,
\end{align}
under which the kinetic terms (\ref{kinetic}) are transformed
so as to cancel the contribution (\ref{var one}):
$\delta_\qq^{(1)}S^{(0)}+\delta_\qq^{(0)}S^{(1)}=0$\,.
Then at the next order we have two contributions, $\delta_\qq^{(1)}S^{(1)}$
and $\delta_\qq^{(0)}S^{(2)}$, which are again nonzero and require to add
\begin{align}
 \delta^{(2)}_\qq \Phi\ =&\ 
\frac{1}{12}[\Phi,[\Phi,\qq\Xi\Psi]]
+\frac{1}{2}\{[\Phi,\Psi],\Xi \qq\Phi\}
+\frac{1}{2}[\Xi[\Phi,\Psi],\qq\Phi]
\nonumber\\
&
+\frac{1}{2}\{\Psi,\Xi\{\eta\Phi,\Xi \qq\Phi\}\}
+\frac{1}{2}\{\Psi,\Xi[\Phi,\Xi \qq\eta\Phi]\}
 -[\Xi[\Phi,\Psi],\Xi \qq\eta\Phi]
\nonumber\\
&
-\frac{1}{2}\qq\Xi[\Phi,\Xi\{\eta\Phi,\Psi\}]
-\frac{1}{2}\qq\Xi[\eta\Phi,\Xi[\Phi,\Psi]],\\
 \delta^{(2)}_\qq \Psi\ =&\ \frac{1}{6}X\eta[\Phi,[\Phi,\qq\Phi]]
 +\frac{1}{2}X\eta[\Phi,\Xi[\qq\Phi,\eta\Phi]]
+\frac{1}{2}X\eta\{\eta\Phi,\Xi[\Phi,\Xi \qq\eta\Phi]\}
\nonumber\\
&
+\frac{1}{2}X\eta[\Phi,\Xi[\eta\Phi,\Xi \qq\eta\Phi]]\,,
\end{align}
to cancel them by $\delta_\qq^{(2)}S^{(0)}$\,:\,
$\delta_\qq^{(2)}S^{(0)}+\delta_\qq^{(1)}S^{(1)}+\delta_\qq^{(0)}S^{(2)}=0$\,.
The procedure is not terminated, so we suppose a full transformation
consistent with these results, and then show that it is in fact a symmetry of 
the complete action.

\subsection{Complete space-time supersymmetry transformation}

Here we suppose that the complete transformation is given by
\begin{subequations} \label{complete transformation}
 \begin{align}
 A_{\delta_\qq}\ =&\ e^\Phi(\qq\Xi(e^{-\Phi}F\Psi e^\Phi))e^{-\Phi} 
 + \{F\Psi,F\Xi A_\qq\},
\label{complete tf ns}\\
\delta_\qq\Psi\ =&\ X\eta F\Xi D_\eta A_\qq\ =\ X\eta F\Xi \qq A_\eta\,,
\label{complete tf r}
\end{align}
\end{subequations}
and show that the complete action (\ref{complete action}) is 
invariant under this transformation. 
From the formula of the general variation of the action (\ref{general variation}),
we have
%
\begin{align}
 \delta_\qq S\ =&\
-\langle e^\Phi(\qq\Xi(e^{-\Phi}F\Psi e^\Phi))e^{-\Phi} ,QA_\eta+(F\Psi)^2\rangle
-\langle \{F\Psi,F\Xi A_\qq\}, QA_\eta+(F\Psi)^2\rangle
\nonumber\\
&\
- \llangle X\eta F\Xi D_\eta A_\qq\,,Y(Q\Psi+X\eta F\Psi)\rrangle\,.
\label{var S}
\end{align}
We calculate each of these three terms, which we denote
(I), (II), and (III), separately.
First, using $(\ref{BPZ homotopy R})$ and the cyclicity of the inner product, 
the second term is calculated as
\begin{equation}
\textrm{(II)} 
=\ 
\langle A_\qq\,, F\Xi[QA_\eta+(F\Psi)^2, F\Psi]\rangle\,.
\label{II}
\end{equation}
For the third term, we find
\begin{align}
\textrm{(III)} =&\
- \llangle \eta F\Xi D_\eta A_\qq\,, Q\Psi+X\eta F\Psi\rrangle 
\nonumber\\
 =&\
- \langle A_\qq\,, D_\eta F\Xi (Q\Psi+X\eta F\Psi) \rangle 
\nonumber\\
 =&\
- \langle A_\qq\,, F (Q\Psi+X\eta F\Psi) \rangle\,,
\label{III}
\end{align}
where we have used $(\ref{BPZ homotopy R})$, $(\ref{important property})$, and
the fact that $X$ is BPZ even with respect to the inner product in the small
Hilbert space, $\llangle XA\,, B\rrangle=\llangle A\,, XB\rrangle$,
and $Q\Psi+X\eta F\Psi$ is in the restricted small Hilbert space.
In order to calculate the first term (I), some consideration is necessary.
%
%
%
In addition to the cyclicity, we need the following relation
for two graded commutative derivations of the string product,
$\mathcal{O}_1$ and $\mathcal{O}_2$
satisfying $\{\mathcal{O}_1\,,\mathcal{O}_2]=0$\,.
\begin{align}
e^{-\Phi}(\mathcal{O}_1A_{\mathcal{O}_2})e^\Phi\
=&\ \mathcal{O}_1\widetilde{A}_{\mathcal{O}_2}+
\widetilde{A}_{\mathcal{O}_1}\widetilde{A}_{\mathcal{O}_2} 
-(-1)^{\mathcal{O}_1\mathcal{O}_2}
\widetilde{A}_{\mathcal{O}_2} \widetilde{A}_{\mathcal{O}_1} 
\nonumber\\
=&\ (-1)^{\mathcal{O}_1\mathcal{O}_2}\mathcal{O}_2\widetilde{A}_{\mathcal{O}_1}\,,
\label{dual relation}
\end{align}
where 
$\widetilde{A}_{\mathcal{O}}$ is 
an analog of the 
left-invariant current:
$\widetilde{A}_{\mathcal{O}}=e^{-\Phi}(\mathcal{O}e^\Phi)$\,.
%
If we use this relation for $(\mathcal{O}_1,\mathcal{O}_2)=(Q,\eta)$, we find
\begin{align}
\textrm{(I)}\ =&\
-\langle \qq\Xi(e^{-\Phi}F\Psi e^\Phi), e^{-\Phi}(QA_\eta+(F\Psi)^2)e^\Phi\rangle
\nonumber\\
=&\ 
\langle \qq\Xi(e^{-\Phi}F\Psi e^\Phi), \eta\widetilde{A}_Q\rangle
-\langle \qq\Xi(e^{-\Phi}F\Psi e^\Phi), (e^{-\Phi}F\Psi e^\Phi)^2\rangle\,.
\label{I-1}
\end{align}
Here the second term vanishes owing to (\ref{BPZ S}) and (\ref{small to large}):
\begin{align}
- \langle \qq\Xi(e^{-\Phi}F\Psi e^\Phi), (e^{-\Phi}F\Psi e^\Phi)^2\rangle\ =&\
\llangle (e^{-\Phi}F\Psi e^\Phi), \{(e^{-\Phi}F\Psi e^\Phi), \qq(e^{-\Phi}F\Psi e^\Phi)\}\rrangle
\nonumber\\
=&\
\frac{2}{3}\Big(
\llangle\qq(e^{-\Phi}F\Psi e^\Phi), (e^{-\Phi}F\Psi e^\Phi)^2\rrangle
\nonumber\\
&\hspace{10mm} 
+ \llangle(e^{-\Phi}F\Psi e^\Phi), \{(e^{-\Phi}F\Psi e^\Phi), \qq(e^{-\Phi}F\Psi e^\Phi)\}\rrangle
\Big)
\nonumber\\
=&\ 0\,.
\end{align}
The first term in (\ref{I-1}) can further be calculated as
\begin{align}
\textrm{(I)} =&\
- \langle \qq(e^{-\Phi}F\Psi e^\Phi), \widetilde{A}_Q\rangle\
=\
\langle F\Psi, e^\Phi(\qq\widetilde{A}_Q)e^{-\Phi}\rangle
\nonumber\\
 =&\
\langle F\Psi, QA_\qq\rangle\
=\ \langle A_\qq, QF\Psi\rangle\,,
\label{I}
\end{align}
where we have used the relation (\ref{dual relation}) with $(\mathcal{O}_1,\mathcal{O}_2)=(Q,\qq)$,
and the identity
\begin{align}
 \eta(e^{-\Phi}F\Psi e^\Phi)\ =&\ e^{-\Phi} (D_\eta F\Psi) e^\Phi\ =\ 0\,.
\end{align}
Summing (\ref{II}), (\ref{III}), and (\ref{I}), 
the variation of the action under the space-time supersymmetry 
transformation finally becomes
\begin{equation}
 \delta_\qq S\ =\ 
\langle A_\qq, \left(QF\Psi - F(Q\Psi+X\eta F\Psi)
+ F\Xi[QA_\eta+(F\Psi)^2, F\Psi]\right)\rangle\,,
\end{equation}
which vanishes
due to the identity (4.89) in Ref.\citen{Kunitomo:2015usa}:
$\delta_\qq S=0$\,. 
Hence the complete action (\ref{complete action}) is invariant 
under the transformation (\ref{complete transformation}).

\section{Algebra of transformation}\label{sec algebra}

Starting from a natural linear transformation (\ref{linearized tf}),
we have constructed the nonlinear 
transformation  (\ref{complete transformation}) as a symmetry of 
the complete action (\ref{complete action}).
If this is in fact space-time supersymmetry, the commutator of two transformations
should satisfy the supersymmetry algebra
\begin{equation}
 [\delta_{\qq_1},\,\delta_{\qq_2}]\ =^{\hspace{-2mm} ?}\ \delta_{p(v_{12})}\,,
\label{susy alg}
\end{equation}
up to the equations of motion (\ref{equations of motion})
and gauge transformation (\ref{full gauge}) generated by some field-dependent parameters,
where $\delta_{p(v_{12})}$ is the space-time translation defined by
\begin{equation}
 \delta_{p(v)}A_\eta\ =\ - p(v)A_\eta\,,
\qquad \delta_{p(v)}\Psi\ =\ - p(v)\Psi\,,
\label{translation}
\end{equation}
with the parameter $v_{12}$ in (\ref{1st quantized alg}).
In this section, we show that the algebra (\ref{susy alg})
is slightly modified, but still the transformation 
(\ref{complete transformation}) can be identified
with space-time supersymmetry.

\subsection{Preparation}

As preparation, note that the relations
\begin{subequations} \label{large small}
 \begin{align}
 \delta A_\eta\ =&\ D_\eta A_\delta\,,
\label{delta eta}\\
 A_\delta\ 
=&\ f\xi_0 \delta A_\eta + D_\eta \Omega_\delta\,,
\label{A delta}
\end{align}
\end{subequations}
hold with $\Omega_\delta=f\xi_0A_\delta$\,,
for general variation of the NS string field $A_\delta$\,.
The former, (\ref{delta eta}), is the case of $(\mathcal{O}_1,\mathcal{O}_2)=(\delta,\eta)$
in (\ref{MC}), and the latter, (\ref{A delta}), is obtained by decomposing $A_\delta$ 
by the projection operators (\ref{proj ns}) and using (\ref{delta eta}). 
These relations (\ref{large small}) show that 
two variations $A_\delta$ and $\delta A_\eta$ 
are in one-to-one correspondence 
up to the $\Omega$-gauge transformation. Since any transformation of the string
field is a special case of the general variation, (\ref{large small}) holds
for any symmetry transformation $\delta_I$\,,
\begin{subequations} \label{delta I} 
\begin{align}
 \delta_I A_\eta\ =&\ D_\eta A_{\delta_I}\,,
\label{delta I 1}\\
 A_{\delta_I}\ =&\ f\xi_0\delta_IA_\eta + D_\eta\Omega_I\,.
\label{delta I 2}
\end{align} 
\end{subequations}
This is the case even for
the commutator of the two transformations $[\delta_I, \delta_J]$\,,
\begin{subequations} \label{delta I delta J}
\begin{align}
 [\delta_I, \delta_J] A_\eta\ =&\ D_\eta A_{[\delta_I, \delta_J]}\,,
\label{delta I delta J 1}\\
 A_{[\delta_I, \delta_J]}\ =&\ 
 f\xi_0[\delta_I, \delta_J] A_\eta + D_\eta\Omega_{IJ}\,,
\label{delta I delta J 2}
\end{align} 
\end{subequations}
with
\begin{align}
\Omega_{IJ}\ =&\ 
-f\xi_0[f\xi_0\delta_I A_\eta,\, f\xi_0\delta_J A_\eta]
\nonumber\\
&\
+\delta_I \Omega_{J}
-[f\xi_0\delta_I A_\eta,\, \Omega_{J}]
-\delta_J \Omega_{I}
+[f\xi_0\delta_J A_\eta,\, \Omega_{I}]
-[\Omega_{I},\, D_\eta\Omega_{J}]\,,
\label{Omega IJ}
\end{align}
which can be shown by explicit calculation using (\ref{gen MC}) and (\ref{f ns})
if we assume (\ref{delta I}) with some field-dependent $\Omega_I$\,.
Therefore if the algebra of the transformation is closed 
on $A_\eta$\,,
\begin{equation}
 [\delta_I,\,\delta_J]A_\eta\ =\ \sum_{K\ne\Omega}\delta_K A_\eta\,,
\label{alg A_eta}
\end{equation}
we have
\begin{equation}
 A_{[\delta_I,\delta_J]}\ =\ \sum_{K\ne\Omega} A_{\delta_K} + D_\eta\Omega_{IJ}\
=\ \sum_K A_{\delta_K}\,,
\label{alg AIJ}
\end{equation}
or equivalently, the algebra is also closed on $e^\Phi$\,:
\begin{equation}
 [\delta_I,\,\delta_J]e^\Phi\ =\ \sum_K\delta_K e^\Phi\,.
\end{equation}
with some field-dependent $\Omega_{IJ}$\,.
Here in (\ref{alg A_eta}) we used that $A_\eta$ is invariant under 
the $\Omega$-gauge transformation,
$A_{\delta_\Omega}=D_\eta\Omega$, as seen from (\ref{delta I 1}).

\subsection{$[\delta_{\qq_1},\delta_{\qq_2}]$}

Now let us explicitly calculate the supersymmetry algebra on 
$A_\eta$ and $\Psi$,
which is easier to calculate than the algebra on the fundamental string fields
$\Phi$ (or $e^\Phi$) and $\Psi$ due to their $\Omega$-gauge invariance and enough 
to know that on the fundamental string fields
as was shown in the previous subsection. From (\ref{complete transformation}) we find  
\begin{subequations} 
 \begin{align}
 A_{\delta_\qq}\ =&\ f\xi_0\delta_\qq A_\eta + D_\eta\Omega_\qq\,,\\
\delta_\qq \Psi\ =&\ X\eta F\Xi\qq A_\eta\,,
\label{susy A_eta}
 \end{align}
\end{subequations}
with
\begin{subequations}
\begin{align}
 \delta_\qq A_\eta\ 
=&\ \qq F\Psi + [F\Psi,F\Xi \qq A_\eta]\
=\ D_\qq F\Psi - [F\Psi, D_\eta F\Xi A_\qq]\,,\\
 \Omega_{\qq}\ =&\ f\xi_0\left(e^\Phi(\qq \Xi(e^{-\Phi}F\Psi e^\Phi))e^{-\Phi}
+\{F\Psi, F\Xi A_{\qq}\}\right)\,.
\end{align}
\end{subequations}
Here we used the relations
\begin{equation}
 D_\eta(e^\Phi A e^{-\Phi})\ =\ e^\Phi(\eta A)e^{-\Phi}\,,\qquad
 \eta(e^{-\Phi} A e^\Phi)\ =\ e^{-\Phi}(D_\eta A)e^\Phi\,,
\end{equation}
which hold for a general string field $A$\,.
%
%
The commutator of two transformations on $\Psi$, 
\begin{equation}
[\delta_{\qq_1}, \delta_{\qq_2}]\,\Psi\
=\
\delta_{\qq_1}(X\eta F\Xi \qq_2 A_\eta) - (1\leftrightarrow 2)\,, 
\end{equation}
which is easier and straightforward,
can be calculated as follows. Using (\ref{variation F}), 
(\ref{homotopy relation}) and (\ref{MC}) with
$(\mathcal{O}_1,\mathcal{O}_2)=(\qq,\eta)$ and $(\delta, \eta)$, we can find
\begin{align}
\delta_{\qq_1}(X\eta F\Xi \qq_2 A_\eta) 
=&\
X\eta F\Xi[\delta_{\qq_1}A_\eta,\,F\Xi\qq_2A_\eta]
+ X\eta F\Xi\qq_2(\delta_{\qq_1}A_\eta)  
\nonumber\\
=&\
X\eta F\Xi D_{\qq_2}(\delta_{\qq_1}A_\eta)
+ X\eta F\Xi [D_\eta F\Xi A_{\qq_2}\,,\delta_{\qq_1}A_\eta]\,.
\end{align}
Then, using $[D_\eta, D_\qq]=0$\,,
\begin{align}
 [\delta_{\qq_1}, \delta_{\qq_2}]\,\Psi\
=&\
\Big(
 X\eta F\Xi D_{\qq_2}D_{\qq_1}F\Psi
- X\eta F\Xi [F\Psi, D_{\qq_2}D_\eta F\Xi A_{\qq_1}]
\nonumber\\
&\
- X\eta F\Xi
[D_\eta F\Xi A_{\qq_2}\,,[F\Psi, D_\eta F\Xi A_{\qq_1}]]
\Big)
- (1\leftrightarrow 2)
\nonumber\\
=&\
- X\eta F\Xi D_{\tilde{p}_{12}}F\Psi
\nonumber\\
&\
+ X\eta F\Xi [F\Psi, 
D_\eta\big(D_{\qq_1}F\Xi A_{\qq_2} - D_{\qq_2}F\Xi A_{\qq_1}
+ [F\Xi A_{\qq_1}, D_\eta F\Xi A_{\qq_2}]\big)]\,,
\label{ss on Psi}
\end{align}
where we have used (\ref{generalized D}) and (\ref{1st quantized alg}), 
and denoted $\tilde{p}(v_{12})=\tilde{p}_{12}$\,.
Comparing with (\ref{gauge tf r}),
we find that the second line has the form of the gauge transformation with
the parameter
\begin{align}
D_\eta\Lambda_{\qq_1\qq_2}\ 
=&\
- D_\eta\Big(D_{\qq_1}F\Xi A_{\qq_2}-D_{\qq_2}F\Xi A_{\qq_1}
+ [ F\Xi A_{\qq_1},\, D_\eta F\Xi A_{\qq_2}]\Big)
\nonumber\\
=&\
- A_{\tilde{p}_{12}} 
+ (\qq_1F\Xi\qq_2-\qq_1F\Xi\qq_1)A_\eta
-[F\Xi\qq_1 A_\eta,\, F\Xi\qq_2A_\eta]\,.
\label{Lambda ss}
\end{align}
The second form can be obtained using (\ref{gen MC}),
and will be used below.

In order to calculate the algebra on $A_\eta$, 
we first calculate the transformation of $F\Psi$ using (\ref{variation F}):
\begin{align}
 \delta_\qq F\Psi\ =&\ 
F\Xi\{\delta_\qq A_\eta, F\Psi\} + F\delta_\qq\Psi
\nonumber\\
=&\
FX\eta F\Xi\qq A_\eta + F\Xi\qq (F\Psi)^2
+ F\Xi[(F\Psi)^2, F\Xi\qq A_\eta]
\nonumber\\
=&\ QF\Xi \qq A_\eta 
+ F\Xi \qq\left(QA_\eta + (F\Psi)^2\right)
+F\Xi[QA_\eta + (F\Psi)^2, F\Xi \qq A_\eta]
\nonumber\\
\cong&\ QF\Xi \qq A_\eta\,, 
\label{susy on F psi} 
\end{align}
where the third equality follows from (\ref{Q and FXi}),
and the symbol $\cong$ denotes an equation 
which holds up to the equations of motion.
Then the commutator of two transformations on $A_\eta$ 
\begin{equation}
 [\delta_{\qq_1}, \delta_{\qq_2}]\,A_\eta\
=\ \delta_{\qq_1}\big(\qq_2 F\Psi+[F\Psi, F\Xi \qq_2 A_\eta]\big) 
- (1\leftrightarrow2)\,,
\end{equation}
can be calculated similarly to that on $\Psi$\,.
Since the first term can be calculated as
\begin{align}
\delta_{\qq_1}\big(\qq_2 F\Psi+[F\Psi, F\Xi \qq_2 A_\eta]\big)
=&\
\qq_2(\delta_{\qq_1}F\Psi)+[(\delta_{\qq_1}F\Psi), F\Xi\qq_2A_\eta]
\nonumber\\
&\
+ [F\Psi, F\Xi D_{\qq_2}(\delta_{\qq_1}A_\eta)]
+[F\Psi, F\Xi[D_\eta F\Xi A_{\qq_2}, (\delta_{\qq_1}A_\eta)]]
\nonumber\\
\cong&\
\qq_2QF\Xi \qq_1 A_\eta + [QF\Xi\qq_1 A_\eta, F\Xi\qq_2 A_\eta]
\nonumber\\
&\
+[F\Psi, F\Xi D_{\qq_2}D_{\qq_1}F\Psi]
- [F\Psi, F\Xi D_{\qq_2}[F\Psi\,, D_\eta F\Xi A_{\qq_1}]]
\nonumber\\
&\
+[F\Psi, F\Xi[D_\eta F\Xi A_{\qq_2}, D_{\qq_1}F\Psi]]
\nonumber\\
&\hspace{10mm}
- [F\Psi, F\Xi[D_\eta F\Xi A_{\qq_2}, [F\Psi\,, D_\eta F\Xi A_{\qq_1}]]]\,,
\end{align}
we find
\begin{align}
 [\delta_{\qq_1}, \delta_{\qq_2}]\,A_\eta\
\cong&\
- Q\Big((\qq_1 F\Xi \qq_2 - \qq_2 F\Xi \qq_1)A_\eta - [F\Xi\qq_1A_\eta,
 F\Xi\qq_2A_\eta]\Big)
\nonumber\\
&\
- [F\Psi, F\Xi[D_{\qq_1}, D_{\qq_2}]F\Psi]
- [F\Psi, F\Xi[F\Psi, D_\eta\Lambda_{\qq_1\qq_2}]]
\nonumber\\
=&\
-QA_{\tilde{p}_{12}}-[F\Psi,\,F\Xi D_{\tilde{p}_{12}}F\Psi]
\nonumber\\
&\
-QD_\eta\Lambda_{\qq_1\qq_2} - [F\Psi,\,F\Xi[F\Psi,\,D_\eta\Lambda_{\qq_1\qq_2}]]\,,
\label{ss on Aeta}
\end{align}
using two expressions in (\ref{Lambda ss}).
From (\ref{ss on Psi}), (\ref{ss on Aeta}) and (\ref{alg AIJ}) we can conclude that
the the commutator of two space-time supersymmetry transformations satisfies the algebra
\begin{equation}
 [\delta_{\qq_1},\delta_{\qq_2}]\ \cong\
\delta_{p(v_{12})} 
+ \delta_{g(\Lambda_{\qq_1\qq_2},\Omega_{\qq_1\qq_2})}
+ \delta_{\tilde{p}(v_{12})}\,,
\label{susy alg2}
\end{equation}
with the gauge parameters given in (\ref{Lambda ss}) and (\ref{Omega IJ}).
The last term absent in (\ref{susy alg}) is a new symmetry 
defined by
\begin{subequations}\label{p tilde}  
\begin{align}
A_{\delta_{\tilde{p}(v)}}\ =&\ A_{p(v)}
-f\xi_0\big(QA_{\tilde{p}(v)} + [F\Psi,\,F\Xi D_{\tilde{p}(v)}F\Psi]\big)\,,
\label{p tilde A}\\
\delta_{\tilde{p}(v)}\,\Psi\ =&\ p(v)\Psi - X\eta F\Xi D_{\tilde{p}(v)}F\Psi\,,
\label{p tilde Psi}
\end{align}
\end{subequations}
where the former is determined so as to induce
\begin{align}
\delta_{\tilde{p}(v)}A_\eta\ =&\ 
 D_\eta\Big( A_{p(v)}- f\xi_0\big(QA_{\tilde{p}(v)} + [F\Psi,\,F\Xi D_{\tilde{p}(v)}F\Psi]\big)\Big)
\nonumber\\
\cong&\ p(v)A_\eta - QA_{\tilde{p}(v)} - [F\Psi,\,F\Xi D_{\tilde{p}(v)}F\Psi]\,.
\end{align}
This extra contribution can be absorbed into the gauge transformation, 
up to the equations of motion, at the linearized level as we will see shortly.

Let us consider the transformation (\ref{p tilde})
at the linearized level:
\begin{subequations} \label{translation tilde}
\begin{align}
	\delta_{\tilde{p}}^{(0)}\Phi\ =&\ p(v)\Phi
	- \xi_0 Q\tilde{p}(v)\Phi\ =\ \big(p(v)- X_0\tilde{p}(v)\big)\Phi +Q(\xi_0\tilde{p}(v)\Phi)\,,
\label{trans tilde ns}\\
	\delta_{\tilde{p}}^{(0)}\Psi\ =&\ \big(p(v)-X\tilde{p}(v)\big)\Psi\,.
\label{trans tilde ramond}
\end{align} 
\end{subequations}
%
%
Thanks to (\ref{p tilde p}), the transformation of $\Phi$ 
(\ref{trans tilde ns}) becomes the form of the gauge transformation 
up to the equation of motion at the linearized level:
\begin{equation}
	\delta_{\tilde{p}}^{(0)}\Phi\ 
=\ 
Q \big((M(v) + \xi_0\tilde{p}(v))\Phi\big)
+ \eta \big(\xi_0 M(v) Q\Phi\big)
+ \xi_0M(v)Q\eta\Phi\,.
\label{trans tilde ns 2}
\end{equation}
We can similarly show that the transformation of $\Psi$ in
(\ref{trans tilde ramond}) can also be written as a gauge transformation 
up to the equation of motion at the linearized level as shown in Appendix~\ref{app B}. 
%
Here we assume that the asymptotic condition\cite{Lehmann:1954rq} 
holds for string field theory
as well as the conventional (particle) field theory.
Then, at least perturbatively, we can identify that the transformation
(\ref{translation tilde}), or (\ref{trans tilde ns 2}) and (\ref{B4})
can be interpreted, with appropriate (finite) renormalization, 
as that of asymptotic string fields.
If we further assume asymptotic completeness, this implies that
the extra transformation (\ref{translation tilde})
acts trivially on the on-shell physical states defined by these asymptotic string fields,
and thus the physical S-matrix.
Thus the supersymmetry algebra is realized on the physical S-matrix, and 
we can identify the transformation (\ref{complete transformation}) with
space-time supersymmetry. 

\subsection{Extra unphysical symmetries}\label{extra symm}

We have shown that the supersymmetry algebra is realized on the physical S-matrix
but this is not the end of the story.
The extra transformation $\delta_{\tilde{p}}$ produces another
extra transformation if we consider the nested commutator 
$[\delta_{\qq_1},[\delta_{\qq_2},\delta_{\qq_3}]]$\,.
The extra contribution comes from the commutator $[\delta_\qq,\delta_{\tilde{p}}]$
which is non-trivial because the first-quantized charges
$\qq$ and $\tilde{p}$ are not commutative: $[\qq,\tilde{p}]\ne0$\,.
In fact, we can show that the algebra
\begin{equation}
 [\delta_\qq,\, \delta_{\tilde{p}}]\ \cong\
\delta_g + \delta_{[\qq,\tilde{p}]}\,,
\label{alg sp}
\end{equation}
holds with the gauge parameters,
\begin{subequations} 
 \begin{align}
 \Lambda_{\qq\tilde{p}}\ =&\
f\xi_0\big(D_{\tilde{p}}f\xi_0D_\qq - D_\qq F\Xi D_{\tilde{p}}\big) F\Psi
- [F\Psi, F\Xi D_{\tilde{p}}F\Xi A_\qq]
\nonumber\\
&\
- [F\Xi A_\qq, F\Xi D_{\tilde{p}}F\Psi]
- D_{\tilde{p}}f\xi_0\{F\Psi, F\Xi A_\qq\}\,,\\
\lambda_{\qq\tilde{p}}\ =&\
X\eta F\Xi D_\eta D_{\tilde{p}} F\Xi A_\qq\,,
\end{align}
\end{subequations}
and $\Omega_{\qq\tilde{p}}$ in (\ref{Omega IJ}).
The new transformation $\delta_{[\qq,\tilde{p}]}$ is
defined by
\begin{subequations} \label{tf sp}
 \begin{align}
A_{\delta_{[\qq,\tilde{p}]}}\ =&\
f\xi_0\Big(Qf\xi_0D_{[\qq,\tilde{p}]}F\Psi
+
  [F\Psi,\,F\Xi\big(QA_{[\qq,\tilde{p}]}+[F\Psi,\,f\xi_0D_{[\qq,\tilde{p}]}F\Psi]\big)]\Big)\,,
\label{tf sp ns}\\
\delta_{[\qq,\tilde{p}]}\Psi\ =&\
X\eta F\Xi\big(
QA_{[\qq,\tilde{p}]}+[F\Psi,\,f\xi_0D_{[\qq,\tilde{p}]}F\Psi]\big)\,,
\label{tf sp ramond}
\end{align}
\end{subequations}
where $[\qq,\tilde{p}]$ denotes the first-quantized charge 
defined by the commutator $[q^\alpha,\tilde{p}^\mu]$ with the parameter $\zeta_{\mu\alpha}$\,,
\begin{equation}
 [\qq,\tilde{p}]\ =\ \zeta_{\mu\alpha} [q^\alpha,\tilde{p}^\mu]\,,
\end{equation}
and in particular $\zeta_{\mu\alpha}=\epsilon_\alpha v_\mu$ on the right-hand side of
(\ref{alg sp}).
This new symmetry is also unphysical in a similar sense to $\delta_{\tilde{p}}$.
At the linearized level,
the transformation (\ref{tf sp}) becomes\footnote{
In this subsection, the symbol $\cong$ denotes an equation that holds 
up to the linearized equations of motion, $Q\eta\Phi=Q\Psi=0$\,. }
\begin{subequations}
  \begin{align}
  \delta_{[\qq,\tilde{p}]}\Phi\ =&\
\xi_0 Q \xi_0 [\qq,\tilde{p}]\Psi\
=\ \xi_0 X_0[\qq,\tilde{p}]\Psi\,,
\label{tf sp ns at linear}\\
\delta_{[\qq,\tilde{p}]}\Psi\ =&\
X\eta\Xi Q[\qq,\tilde{p}]\Phi\
\cong\ XQ[\qq,\tilde{p}]\Phi\,,
\label{tf sp ramond at linear}  %
 \end{align}
\end{subequations}
where we have used the fact that $\qq$, $\tilde{p}$, and thus $[\qq, \tilde{p}]$
are commutative with $Q$ and $\eta$\,.
If we note that $[\qq,p]=0$\, and  
\begin{equation}
 [\qq,X_0]\ =\ [\qq,\{Q,\xi_0\}]\ =\
\{Q,[\qq,\xi_0]\}+\{\xi_0,[\qq,Q]\}\ =\ 0\,,
\end{equation}
the transformation of $\Phi$\,, (\ref{tf sp ns}), can further be rewritten 
in the form of a linearized gauge transformation:
\begin{align}
  \delta_{[\qq,\tilde{p}]}\Phi\ 
=&\ - \xi_0[\qq, (p-X_0\tilde{p})]\,\Psi\
=\ -\xi_0[\qq, \{Q, M\}]\,\Psi
\nonumber\\
\cong&\ -\xi_0 Q[\qq, M]\,\Psi
\nonumber\\
=&\ Q(\xi_0[\qq, M]\Psi)-\eta(\xi_0X_0[\qq,M]\Psi)\,.
\end{align}
Similarly the transformation of $\Psi$ (\ref{tf sp ramond})
can also be written as
\begin{align}
  \delta_{[\qq,\tilde{p}]}\Psi\ 
\cong&\
X\eta\xi_0Q[\qq,\tilde{p}]\Phi
\nonumber\\
=&\
Q(X\eta\xi_0[\qq,\tilde{p}]\Phi)
+ X\eta X_0[\qq,\tilde{p}]\Phi
\nonumber\\
\cong&\
Q(X\eta\xi_0[\qq,\tilde{p}]\Phi
+ X\eta[\qq,M]\Phi)\,.
\end{align}
It should be noted that the gauge parameter
in this form,
$\lambda_{\qq\tilde{p}}=X\eta\xi_0[\qq,\tilde{p}]\Phi
+ X\eta[\qq,M]\Phi$\,,
is in the restricted small Hilbert space:
$\eta\lambda_{\qq\tilde{p}}=0$ and
$XY\lambda_{\qq\tilde{p}}=\lambda_{\qq\tilde{p}}$\,.

In addition, a further extra transformation is produced
by considering the commutator between $\delta_{\tilde{p}_1}$
and $\delta_{\tilde{p}_2}$, 
and this sequence of extra transformations does not terminate
as long as the nested commutators,
$[\oo,[\oo,\oo]]$, $[\oo,[\oo,[\oo,\oo]]]$, $\cdots$, 
with $\mathcal{O}=$ $\qq$ or $\tilde{p}$\,, do not vanish.
%
This complicates the structure of the algebra, but
we can similarly show that all of these extra transformations act
trivially on the physical S-matrix, as shown in Appendix~\ref{app B}.

\section{Summary and discussion}

In this paper, we have explicitly constructed a space-time supersymmetry transformation
of the WZW-like open superstring field theory in flat ten-dimensional 
space-time. Under the GSO projections, we have extended a linear
transformation expected from space-time supersymmetry 
in the first-quantized theory
to a nonlinear transformation so as to
be a symmetry of the complete action (\ref{complete action}).
We have also shown that the transformation satisfies the supersymmetry algebra 
up to gauge transformation, the equations of motion and a transformation $\delta_{\tilde{p}}$
acting trivially on the asymptotic physical states defined by the asymptotic string fields.
This unphysical transformation produces a series of transformations
$\delta_{[\qq,\tilde{p}]},\, \delta_{[\tilde{p}\tilde{p}]},\,\cdots$
by taking commutators with $\delta_\qq$ or $\delta_{\tilde{p}}$ repeatedly.
%
All of these symmetries also act trivially on the asymptotic physical states,
and thus are unphysical, but it is interesting
to clarify their complete structure, which is nontrivial
in the total Hilbert space including unphysical 
degrees of freedom.

In any case, except for such an unphysical complexity, 
we have now understood how space-time supersymmetry is realized 
in superstring field theory, and therefore are ready to study various
consequences of space-time supersymmetry\cite{Kishimoto:2005bs}\tocite{Sen:2015uoa}
on a firm basis.
We have to (re)analyze them precisely using the techniques developed 
in conventional quantum field theory.\footnote{For such analyses of
superstring field theory, see, for example, 
Refs. \citen{Pius:2016jsl}-\citen{Ishibashi:2016jno}.} 
We hope to report on them in the near future.

\section*{Acknowledgements}

The author would like to give special thanks to Ted~Erler for helpful discussion,
that was essential for clarifying the structure of the supersymmetry algebra 
in the large Hilbert space. 
The main part of the work was completed at the workshop on 
``String Field Theory and Related Aspects VIII''
held at ICTP, SAIFR in S\~ao Paulo, Brazil. The author also thanks to the organizers,
particularly Nathan Berkovits, for their hospitality and for 
providing a stimulating atmosphere.

\appendix

\section{Spinor conventions and Ramond ground states}\label{convention}

In this paper, although it is mostly implicit,
we adopt the chiral representation for $SO(1,9)$ gamma matrices $\Gamma^\mu$,
in which $\Gamma^\mu$ is given by
\begin{equation}
 \Gamma^\mu\ =\ 
\begin{pmatrix}
 0 & (\gamma^\mu)_{\alpha\dot{\beta}}\\
(\bar{\gamma}^\mu)^{\dot{\alpha}\beta} & 0
\end{pmatrix}\,,
\end{equation}
where $\gamma^\mu$ and $\bar{\gamma}^\mu$ satisfy
\begin{equation}
 {(\gamma^\mu\bar{\gamma}^\nu + \gamma^\nu\bar{\gamma}^\mu)_\alpha}^\beta
=\ 2\eta^{\mu\nu}{\delta_\alpha}^\beta\,,\qquad
 {(\bar{\gamma}^\mu\gamma^\nu + \bar{\gamma}^\nu\gamma^\mu)^{\dot{\alpha}}}_{\dot{\beta}}
=\ 2\eta^{\mu\nu}{\delta^{\dot{\alpha}}}_{\dot{\beta}}\,.
\end{equation}
The charge conjugation matrix $\mathcal{C}$ satisfies the relations
\begin{equation}
 (\Gamma^\mu)^T\ =\ -\mathcal{C}\Gamma^\mu\mathcal{C}^{-1},\qquad
\mathcal{C}^T\ =\ -\mathcal{C}\,,
\end{equation}
and is given in the chiral representation by
\begin{equation}
 \mathcal{C}\ =\
\begin{pmatrix}
 0 & {C^\alpha}_{\dot{\beta}}\\
 -{(C^T)_{\dot{\alpha}}}^\beta & 0
\end{pmatrix}\,.
\end{equation}
The matrices $\mathcal{C}\Gamma^\mu$ are symmetric, or equivalently
\begin{equation}
 (C\bar{\gamma}^\mu)^{\alpha\beta}\ =\  (C\bar{\gamma}^\mu)^{\beta\alpha}\,,\qquad
 (C^T\gamma^\mu)_{\dot{\alpha}\dot{\beta}}\ =\  (C^T\gamma^\mu)_{\dot{\beta}\dot{\alpha}}\,.
\end{equation}


The world-sheet fermion $\psi^\mu(z)$ in the Ramond sector has zero-modes that satisfy
the $SO(1,9)$ Clifford algebra
\begin{equation}
 \{\psi^\mu_0, \psi^\nu_0\}\ =\ 0\,.
\end{equation}
The degenerate ground states therefore become the space-time spinor, on which
$\psi^\mu_0$ act as space-time gamma matrices. 
We summarize here the related convention.
We denote the ground state spinor as 
$\left(\begin{matrix}|{}^\alpha\rangle\\ |{}_{\dot{\alpha}}\rangle\end{matrix} \right)$\,,
on which $\psi^\mu_0$ acts as
\begin{equation}
\psi^\mu_0|{}^\alpha\rangle\ =\ 
|{}_{\dot{\alpha}}\rangle\frac{1}{\sqrt{2}}(\bar{\gamma}^\mu)^{\dot{\alpha}\alpha}\,,\qquad
\psi^\mu_0|{}_{\dot{\alpha}}\rangle\ =\
|{}^\alpha\rangle\frac{1}{\sqrt{2}}(\gamma^\mu)_{\alpha\dot{\alpha}}\,.
\end{equation}
Then $\hat{\Gamma}_{11}$ defined by (\ref{gamma11}) acts on the ground states as
\begin{equation}
 \hat{\Gamma}_{11}|{}^\alpha\rangle\ =\ |{}^\alpha\rangle\,,\qquad
 \hat{\Gamma}_{11}|{}_{\dot{\alpha}}\rangle\ =\ -|{}_{\dot{\alpha}}\rangle\,,
\end{equation}
by which the definition of the GSO projection (\ref{GSO Ramond}) is supplemented.
Similarly, the BPZ conjugate of the ground state spinor
$(\langle{}^\alpha|,\langle{}_{\dot{\alpha}}|)$
satisfies
\begin{equation}
 \langle{}^\alpha|\psi^\mu_0\ =\ 
\frac{i}{\sqrt{2}}(\bar{\gamma}^\mu)^{\dot{\alpha}\alpha}\langle{}_{\dot{\alpha}}|\,,\qquad
 \langle{}_{\dot{\alpha}}|\psi^\mu_0\ =\
- \frac{i}{\sqrt{2}}(\gamma^\mu)_{\alpha\dot{\alpha}}\langle{}^\alpha|\,,
\end{equation}
with the normalization
\begin{equation}
 \langle{}^\alpha|{}_{\dot{\alpha}}\rangle\ =\ {C^\alpha}_{\dot{\alpha}}\,,\qquad
 \langle{}_{\dot{\alpha}}|{}^\alpha\rangle\ =\ 
-i{C^\alpha}_{\dot{\alpha}}\,.
\end{equation}
The nontrivial matrix elements of $\psi^\mu_0$ are then given by
\begin{equation}
 \langle{}^\alpha|\psi^\mu_0|{}^\beta\rangle\ 
=\ \frac{1}{\sqrt{2}}(C\bar{\gamma}^\mu)^{\alpha\beta}\,,\qquad
 \langle{}_{\dot{\alpha}}|\psi^\mu_0|{}_{\dot{\beta}}\rangle\ =\
-\frac{i}{\sqrt{2}}(C^T\gamma^\mu)_{\dot{\alpha}\dot{\beta}}\,.
\end{equation}

%
%
%
%

\section{Triviality of the extra unphysical symmetries at the linearized level}\label{app B}

First, in order to show the triviality of (\ref{trans tilde ramond}), 
it is useful to introduce the local inverse picture-changing operator  
\begin{equation}
	Y(z_0)\ =\ -c(z_0)\delta'(\gamma(z_0))\,,
\end{equation}
which also satisfies
\begin{equation}
	XY(z_0)X\ =\ X\,,\label{xyx local}
\end{equation}
and in addition is commutative with $Q$\,: $[Q, Y(z_0)]=0$. 
The point $z_0$ can be chosen to be any point on the string, for example, the midpoint $z_0=i$\,.
Due to (\ref{xyx local}), 
we can define another projection operator $XY(z_0)$ that is commutative with $Q$\,,
and acts identically with $XY$ in the restricted small Hilbert space:
\begin{equation}
[Q, XY(z_0)]\ =\ 0\,,
\end{equation}
and if $XY\Psi=\Psi$ then
\begin{equation}
 XY(z_0)\Psi\ =\ XY(z_0)XY\Psi\ =\ XY\Psi\,.
\end{equation}
%
%

Using this projection operator, the linearized transformation 
(\ref{trans tilde ramond}) can be written as
the a linearized gauge transformation,
\begin{align}
\delta_{\tilde{p}}^{(0)}\Psi\ 
 =&\ 
XY(z_0)\left(p(v) - X \tilde{p}(v)\right)\Psi\ 
=\ XY(z_0)\{Q, \tilde{M}(v)\}\,\Psi\
\nonumber\\
\cong&\ Q(XY(z_0)\tilde{M}(v)\Psi)\,, 
\label{B4}
\end{align}
up to the linearized equation of motion, $Q\Psi=0$\,, with
\begin{equation}
  \tilde{M}(v)\ =\  v^\mu
\oint\frac{dz}{2\pi i}(\xi(z)-\Xi)\psi_\mu(z)e^{-\phi(z)}\,.
\end{equation}
We can see that the gauge parameter in (\ref{B4}),
\begin{equation}
 \lambda_{\tilde{p}}\ =\ XY(z_0)\tilde{M}(v)\Psi\,,
\end{equation}
is in the restricted small Hilbert space,
\begin{equation}
\eta  \lambda_{\tilde{p}}\ =\ 0\,,\qquad
XY  \lambda_{\tilde{p}}\ =\  \lambda_{\tilde{p}}\,,
\end{equation}
if we note that $\{\eta, \tilde{M}\}=0$\,.

As was mentioned in section \ref{extra symm},
the commutator $[\delta_{\tilde{p}_1},\delta_{\tilde{p}_2}]$
produces another unphysical transformation 
$\delta_{[\tilde{p},\tilde{p}]}$\,:
\begin{equation}
 [\delta_{\tilde{p}_1},\, \delta_{\tilde{p}_2}]\ \cong\
\delta_g + \delta_{[\tilde{p},\tilde{p}]_{12}}\,,
\label{alg pp}
\end{equation}
where the field-dependent parameters are given by
\begin{align}
 \Lambda_{\tilde{p}_1\tilde{p}_2}\ =&\
f\xi_0\Big((D_{\tilde{p}_1}f\xi_0 D_{\tilde{p}_2} 
- D_{\tilde{p}_2}f\xi_0 D_{\tilde{p}_1})A_Q
+D_{\tilde{p}_1}f\xi_0[F\Psi, F\Xi D_{\tilde{p}_2}F\Psi]
\nonumber\\
&\
-D_{\tilde{p}_2}f\xi_0[F\Psi, F\Xi D_{\tilde{p}_1}F\Psi]
+\{F\Psi, F\Xi(D_{\tilde{p}_1}F\Xi D_{\tilde{p}_2}
- D_{\tilde{p}_2}F\Xi D_{\tilde{p}_1})F\Psi\}
\nonumber\\
&\
-[F\Xi D_{\tilde{p}_2}F\Psi, F\Xi D_{\tilde{p}_2}F\Psi]\Big)\,,\\
\lambda_{\tilde{p}_1\tilde{p}_2}\ =&\
-X\eta F\Xi(D_{\tilde{p}_1}F\Xi D_{\tilde{p}_2} 
- D_{\tilde{p}_2}F\Xi D_{\tilde{p}_1})F\Psi\,,
\end{align}
and $\Omega_{\tilde{p}_1\tilde{p}_2}$ in (\ref{Omega IJ}).
The unphysical transformation $\delta_{[\tilde{p},\tilde{p}]}$
is defined by
\begin{subequations} \label{tf pp} 
\begin{align}
A_{\delta_{[\tilde{p},\tilde{p}]}}\ =&\
- f\xi_0\Bigg(Qf\xi_0\big(
QA_{[\tilde{p},\tilde{p}]}+[F\Psi, \,F\Xi D_{[\tilde{p},\tilde{p}]}]F\Psi]\big)
\nonumber\\
&\
+ [F\Psi,\,F\Xi\Big(
QF\Xi D_{[\tilde{p},\tilde{p}]}F\Psi
+[F\Psi,f\xi_0\big(
QA_{[\tilde{p},\tilde{p}]}+[F\Psi, F\Xi D_{[\tilde{p},\tilde{p}]}F\Psi]\big)]
\Big)]\Bigg)\,,
\label{tf pp ns}\\
\delta_{[\tilde{p},\tilde{p}]}\Psi\ =&\
-X\eta F\Xi\Big(
QF\Xi D_{[\tilde{p},\tilde{p}]}F\Psi
+[F\Psi,\,f\xi_0\big(
QA_{[\tilde{p},\tilde{p}]}+[F\Psi, \,F\Xi D_{[\tilde{p},\tilde{p}]}F\Psi]\big)]
\Big)\,.
\label{tf pp ramond}
\end{align}
\end{subequations}
The first-quantized charge $[\tilde{p},\tilde{p}]$ is defined by
\begin{equation}
 [\tilde{p},\tilde{p}]\ =\ w_{\mu\nu}[\tilde{p}^\mu,\tilde{p}^\nu]\,,
\end{equation}
with the parameter $w_{\mu\nu}\,(=-w_{\nu\mu})$, 
and $[\tilde{p},\tilde{p}]_{12}
=[\tilde{p},\tilde{p}](w_{12}=(v_1v_2-v_2v_1)/2)$\, in (\ref{alg pp}).
At the linearized level, the transformation (\ref{tf pp}) becomes
 \begin{align}
  \delta_{[\tilde{p},\tilde{p}]}\ \Phi\ =&\
-\xi_0Q\xi_0Q[\tilde{p},\tilde{p}]\Phi\
=\ -\xi_0QX_0[\tilde{p},\tilde{p}]\Phi\,,\\
  \delta_{[\tilde{p},\tilde{p}]}\ \Psi\ =&\
-X\eta\Xi Q\Xi[\tilde{p},\tilde{p}]\Psi\
=\ -X\eta\Xi X[\tilde{p},\tilde{p}]\Psi\,,
 \end{align}
and can further be rewritten in the form of
a linearized gauge transformation:
\begin{align}
  \delta_{[\tilde{p},\tilde{p}]}\ \Phi\ =&\
\xi_0 Q [\tilde{p},\{Q, M\}]\Phi\
=\
\xi_0 Q [\tilde{p}, M]Q\Phi
\nonumber\\
\cong&\
-Q(\xi_0[\tilde{p}, M]Q\Phi)
+ \eta(\xi_0X_0[\tilde{p}, M]Q\Phi)\,,
\end{align}
and
\begin{align}
 \delta_{[\tilde{p},\tilde{p}]}\ \Psi\ 
=&\
X\eta\Xi[\tilde{p},\{Q, \tilde{M}\}]\Psi\
\cong\
X\eta\Xi Q[\tilde{p}, \tilde{M}]\Psi
\nonumber\\
=&\
Q(X\eta\Xi[\tilde{p},\tilde{M}]\Psi)\,,
\end{align}
up to the linearized equations of motion.
The parameter $\lambda_{\tilde{p}\tilde{p}}=X\eta\Xi[\tilde{p},\tilde{M}]\Psi$
is in the restricted small Hilbert space:
$\eta\lambda_{\tilde{p}\tilde{p}}=0$\, and 
$XY\lambda_{\tilde{p}\tilde{p}}=\lambda_{\tilde{p}\tilde{p}}$\,.

Finally we show that all the extra symmetries obtained from the repeated commutators of
$\delta_\qq$'s and $\delta_{\tilde{p}}$'s act trivially on the physical states
defined by the asymptotic string fields. 
For this purpose, it is enough to consider the transformations of $\eta\Phi$ and $\Psi$ 
at the linearized level for a similar reason to that discussed in Section~\ref{sec algebra}.
Using the linearized form of (\ref{large small}) for general variation, 
\begin{equation}
 \delta\Phi\ =\ \xi_0\delta\eta\Phi + \eta(\xi_0\delta\Phi)\,,
\end{equation}
we can show that if the transformation of $\eta\Phi$ has the form of
a gauge transformation, $\delta\eta\Phi= - Q\eta\Lambda$, 
with some field-dependent parameter $\Lambda$\,, then the transformation of $\Phi$
also has the form of a gauge transformation:
\begin{align}
\delta\Phi\ =&\ - \xi_0 Q\eta\Lambda + \eta\Omega
\nonumber\\
=&\ Q\Lambda + \eta(\Omega-\xi_0Q\Lambda)\,,   
\end{align}
with some field-dependent $\Omega$\,.

Starting from the linearized transformations
\begin{alignat}{3}
 \delta_\qq\eta\Phi\ =&\ \qq\Psi\,,\qquad & \delta_\qq\Psi\ =&\ X\qq\eta\Phi\,,\\
 \delta_{\tilde{p}}\eta\Phi\ =&\ (p-X_0\tilde{p})\eta\Phi\,,\qquad&
 \delta_{\tilde{p}}\Psi\ =&\ (p-X\tilde{p})\Psi\,,
\end{alignat}
extra symmetries can be read from repeated commutators,
$[\delta_{\mathcal{O}_1},[\delta_{\mathcal{O}_2},\cdots,
[\delta_{\mathcal{O}_n},\delta_{\tilde{p}}]\cdots]]$\,, where $\mathcal{O}_i=\qq$
or $\tilde{p}$\,. 
For example, we can read $\delta_{[\qq,\tilde{p}]}$ from $[\delta_\qq,\delta_{\tilde{p}}]$,
\begin{align}
 [\delta_\qq, \delta_{\tilde{p}}]\,\eta\Phi\ =&\
(p-X_0\tilde{p})\qq\Psi - \qq (p-X\tilde{p})\Psi
\nonumber\\
=&\
- X_0\tilde{p}\qq\Psi + \qq X\tilde{p}\Psi
\nonumber\\
\cong&\
-X_0\tilde{p}\qq\Psi + Q\qq\{\xi_0,\eta\}\Xi\tilde{p}\Psi
\nonumber\\
=&\
[\qq, X_0\tilde{p}]\Psi + Q\eta(\qq\xi_0\Xi\tilde{p}\Psi)
\nonumber\\
=&\ - [\qq, (p-X_0\tilde{p})]\Psi + Q\eta(\qq\xi_0\Xi\tilde{p}\Psi)\,,
\label{sp on etaphi}
\end{align}
and
\begin{align}
 [\delta_\qq, \delta_{\tilde{p}}]\,\Psi\ =&\
(p-X\tilde{p})X\qq\eta\Phi - X\qq(p-X_0\tilde{p})\eta\Phi
\nonumber\\
=&\ -X\tilde{p}X\qq\eta\Phi + X\qq X_0\tilde{p}\eta\Phi
\nonumber\\
\cong&\
-QX\{\xi_0,\eta\}\tilde{p}\Xi\qq\eta\Phi + X\qq X_0\tilde{p}\eta\Phi
\nonumber\\
=&\
X[\qq, X_0\tilde{p}]\eta\Phi - Q\eta(X\xi_0\tilde{p}\Xi\qq\eta\Phi)
\nonumber\\
=&\
-X[\qq, (p-X_0\tilde{p})]\eta\Phi - Q\eta(X\xi_0\tilde{p}\Xi\qq\eta\Phi)\,,
\label{sp on psi}
\end{align}
as
\begin{align}
 \delta_{[\qq,\tilde{p}]}\eta\Phi\ =&\
- [\qq, (p-X_0\tilde{p})]\Psi\,,\\
 \delta_{[\qq,\tilde{p}]}\Psi\ =&\
- X[\qq, (p-X_0\tilde{p})]\eta\Phi\,,
\end{align}
up to the equations of motion and gauge transformation.
Similarly we can find that general extra symmetries have the form
\begin{subequations}\label{general odd}
\begin{align}
 \delta_{[\mathcal{O}_1,[\mathcal{O}_2,\cdots,[\mathcal{O}_{2k+l-1},\tilde{p}]]]}\,\eta\Phi\ =&\
- (-1)^l(X_0)^{k+l-1}[\mathcal{O}_1,[\mathcal{O}_2,\cdots,[\mathcal{O}_{2k+l-1},(p-X_0\tilde{p})]]]\,\Psi\,,\\
 \delta_{[\mathcal{O}_1,[\mathcal{O}_2,\cdots,[\mathcal{O}_{2k+l-1},\tilde{p}]]]}\,\Psi\ =&\
- (-1)^l(X)^{k+l}[\mathcal{O}_1,[\mathcal{O}_2,\cdots,[\mathcal{O}_{2k+l-1},(p-X_0\tilde{p})]]]\,\eta\Phi\,,
\end{align}
\end{subequations}
or
\begin{subequations}\label{general even}
\begin{align}
 \delta_{[\mathcal{O}_1,[\mathcal{O}_2,\cdots,[\mathcal{O}_{2k+l},\tilde{p}]]]}\,\eta\Phi\ =&\
(-1)^l(X_0)^{k+l}[\mathcal{O}_1,[\mathcal{O}_2,\cdots,[\mathcal{O}_{2k+l},(p-X_0\tilde{p})]]]\,\eta\Phi\,,\\
 \delta_{[\mathcal{O}_1,[\mathcal{O}_2,\cdots,[\mathcal{O}_{2k+l},\tilde{p}]]]}\,\Psi\ =&\
(-1)^l(X)^{k+l}[\mathcal{O}_1,[\mathcal{O}_2,\cdots,[\mathcal{O}_{2k+l},(p-X_0\tilde{p})]]]\,\Psi\,,
\end{align}
\end{subequations}
with $k=1,2,\cdots$\, and $l=0,1,\cdots$\,, up to the equations of motion and gauge transformation. 
Here $2k-1$ $(l)$ of the $\mathcal{O}$'s are $\qq$ ($\tilde{p}$) in (\ref{general odd})
and $2k$ $(l)$ of the $\mathcal{O}$'s are $\qq$ ($\tilde{p}$) in (\ref{general even}).
All the picture-changing operators, except for the last one, can be put together in front of the right-hand side
with aligning $X_0$ or $X$, which is always possible in a similar way to (\ref{sp on etaphi}) or (\ref{sp on psi}).
If an $X$ is in front of some $\mathcal{O}_{i_0}$\,, we can move it to the top, for example,
\begin{align}
& (X_0)^p[\mathcal{O}_1,[\mathcal{O}_2,\cdots,
[X\mathcal{O}_{i_0},\cdots,[\mathcal{O}_n,(p-X_0\tilde{p})]]]]\eta\Phi
\nonumber\\
&\hspace{20mm}
\cong\ Q\{\xi_0,\eta\}(X_0)^p[\mathcal{O}_1,[\mathcal{O}_2,\cdots,
[\Xi\mathcal{O}_{i_0},\cdots,[\mathcal{O}_n,(p-X_0\tilde{p})]]]]\eta\Phi\
\nonumber\\
&\hspace{20mm}
=\ Q\xi_0(X_0)^p[\mathcal{O}_1,[\mathcal{O}_2,\cdots,
[\mathcal{O}_{i_0},\cdots,[\mathcal{O}_n,(p-X_0\tilde{p})]]]]\eta\Phi\
\nonumber\\
&\hspace{25mm}
+Q\eta(\xi_0(X_0)^p[\mathcal{O}_1,[\mathcal{O}_2,\cdots,
[\mathcal{O}_{i_0},\cdots,[\mathcal{O}_n,(p-X_0\tilde{p})]]]]\eta\Phi)\,.
\nonumber\\
&\hspace{20mm}
\cong\ (X_0)^{p+1}[\mathcal{O}_1,[\mathcal{O}_2,\cdots,
[\mathcal{O}_{i_0},\cdots,[\mathcal{O}_n,(p-X_0\tilde{p})]]]]\eta\Phi\
\nonumber\\
&\hspace{25mm}
+Q\eta(\xi_0(X_0)^p[\mathcal{O}_1,[\mathcal{O}_2,\cdots,
[\mathcal{O}_{i_0},\cdots,[\mathcal{O}_n,(p-X_0\tilde{p})]]]]\eta\Phi)\,.
\end{align}
Using (\ref{p tilde p}), it is easy to show that the transformations (\ref{general odd})
or (\ref{general even}) can further be written in the form of a gauge transformation as
\begin{subequations}
\begin{align}
 \delta_{[\mathcal{O}_1,[\mathcal{O}_2,\cdots,[\mathcal{O}_{2k+l-1},\tilde{p}]]]}\eta\Phi
\cong& -(-1)^l Q\eta( (X_0)^{k+l-1}\xi_0[\mathcal{O}_1,[\mathcal{O}_2,\cdots,[\mathcal{O}_{2k+l-1}, M]\cdots]]\Psi),\\
 \delta_{[\mathcal{O}_1,[\mathcal{O}_2,\cdots,[\mathcal{O}_{2k+l-1},\tilde{p}]]]}\,\Psi
\cong& -(-1)^l Q((X)^{k+l}\eta\xi_0[\mathcal{O}_1,[\mathcal{O}_2,\cdots,[\mathcal{O}_{2k+l-1}, M]\cdots]]\,\eta\Phi)\,,
\end{align}
\end{subequations}
or
\begin{subequations}
\begin{align}
 \delta_{[\mathcal{O}_1,[\mathcal{O}_2,\cdots,[\mathcal{O}_{2k+l},\tilde{p}]]]}\,\eta\Phi\ 
\cong&\ (-1)^l Q \eta ((X_0)^{k+l}\xi_0[\mathcal{O}_1,[\mathcal{O}_2,\cdots,[\mathcal{O}_{2k+l}, M]\cdots]]\,\eta\Phi)\,,\\
 \delta_{[\mathcal{O}_1,[\mathcal{O}_2,\cdots,[\mathcal{O}_{2k+l},\tilde{p}]]]}\,\Psi\ 
\cong&\ (-1)^l Q ( (X)^{k+l}\eta\xi_0[\mathcal{O}_1,[\mathcal{O}_2,\cdots,[\mathcal{O}_{2k+l}, M]\cdots]]\,\Psi)\,,
\end{align}
\end{subequations}
respectively.
Hence all the extra symmetries obtained as repeated commutators
of $\delta_\qq$'s and $\delta_{\tilde{p}}$'s act trivially on the on-shell physical
states, and thus the physical S-matrix, defined by the asymptotic string fields.


\begin{thebibliography}{99}


\bibitem{Berkovits:1995ab}
  N.~Berkovits,
  ``SuperPoincare invariant superstring field theory,''
  Nucl.\ Phys.\ B {\bf 450} (1995) 90
   [Erratum-ibid.\ B {\bf 459} (1996) 439]
  [hep-th/9503099].

\bibitem{Erler:2013xta} 
  T.~Erler, S.~Konopka and I.~Sachs,
 ``Resolving Witten's superstring field theory,''
  JHEP {\bf 1404}, 150 (2014)
  doi:10.1007/JHEP04(2014)150
  [arXiv:1312.2948 [hep-th]].

\bibitem{Kunitomo:2015usa} 
  H.~Kunitomo and Y.~Okawa,
  ``Complete action for open superstring field theory,''
  PTEP {\bf 2016}, no. 2, 023B01 (2016)
  doi:10.1093/ptep/ptv189
  [arXiv:1508.00366 [hep-th]].


\bibitem{Erler:2016ybs} 
  T.~Erler, Y.~Okawa and T.~Takezaki,
  ``Complete Action for Open Superstring Field Theory with Cyclic $A_\infty$ Structure,''
  arXiv:1602.02582 [hep-th].

\bibitem{Erler:2016rxg} 
  T.~Erler,
  ``Supersymmetry in Open Superstring Field Theory,''
  arXiv:1610.03251 [hep-th].


\bibitem{Friedan:1985ge}
  D.~Friedan, E.~J.~Martinec and S.~H.~Shenker,
  ``Conformal Invariance, Supersymmetry and String Theory,''
  Nucl.\ Phys.\ B {\bf 271} (1986) 93.

\bibitem{Witten:1986qs} 
  E.~Witten,
  ``Interacting Field Theory of Open Superstrings,''
  Nucl.\ Phys.\ B {\bf 276}, 291 (1986).
  doi:10.1016/0550-3213(86)90298-1


\bibitem{Terao:1985rw} 
  H.~Terao and S.~Uehara,
  ``Covariant Second Quantization of Free Superstring,''
  Phys.\ Lett.\ B {\bf 168}, 70 (1986).
  doi:10.1016/0370-2693(86)91462-0


\bibitem{Gliozzi:1976qd} 
  F.~Gliozzi, J.~Scherk and D.~I.~Olive,
  ``Supersymmetry, Supergravity Theories and the Dual Spinor Model,''
  Nucl.\ Phys.\ B {\bf 122}, 253 (1977).
  doi:10.1016/0550-3213(77)90206-1

\bibitem{Lehmann:1954rq} 
  H.~Lehmann, K.~Symanzik and W.~Zimmermann,
  Nuovo Cim.\  {\bf 1}, 205 (1955).
  doi:10.1007/BF02731765

\bibitem{Kishimoto:2005bs} 
  I.~Kishimoto and T.~Takahashi,
  ``Marginal deformations and classical solutions in open superstring field theory,''
  JHEP {\bf 0511}, 051 (2005)
  doi:10.1088/1126-6708/2005/11/051
  [hep-th/0506240].

\bibitem{Erler:2013wda} 
  T.~Erler,
  ``Analytic solution for tachyon condensation in Berkovits\rq\ open superstring field theory,''
  JHEP {\bf 1311}, 007 (2013)
  doi:10.1007/JHEP11(2013)007
  [arXiv:1308.4400 [hep-th]].

\bibitem{Berkovits:2014rpa} 
  N.~Berkovits and E.~Witten,
  ``Supersymmetry Breaking Effects using the Pure Spinor Formalism of the Superstring,''
  JHEP {\bf 1406}, 127 (2014)
  doi:10.1007/JHEP06(2014)127
  [arXiv:1404.5346 [hep-th]].

\bibitem{Sen:2015uoa} 
  A.~Sen,
  ``Supersymmetry Restoration in Superstring Perturbation Theory,''
  JHEP {\bf 1512}, 075 (2015)
  doi:10.1007/JHEP12(2015)075
  [arXiv:1508.02481 [hep-th]].





\bibitem{Pius:2016jsl} 
  R.~Pius and A.~Sen,
 ``Cutkosky Rules for Superstring Field Theory,''
  JHEP {\bf 1610}, 024 (2016)
  doi:10.1007/JHEP10(2016)024
  [arXiv:1604.01783 [hep-th]].

\bibitem{Sen:2016bwe} 
  A.~Sen,
 ``Reality of Superstring Field Theory Action,''
  JHEP {\bf 1611}, 014 (2016)
  doi:10.1007/JHEP11(2016)014
  [arXiv:1606.03455 [hep-th]].

\bibitem{Sen:2016uzq} 
  A.~Sen,
  ``Unitarity of Superstring Field Theory,''
  arXiv:1607.08244 [hep-th].

\bibitem{Sen:2016qap} 
  A.~Sen,
 ``Wilsonian Effective Action of Superstring Theory,''
  arXiv:1609.00459 [hep-th].

\bibitem{Sen:2016ubf} 
  A.~Sen,
 ``Equivalence of Two Contour Prescriptions in Superstring Perturbation Theory,''
  arXiv:1610.00443 [hep-th].


\bibitem{Ishibashi:2016bno} 
  N.~Ishibashi,
  ``Light-cone gauge superstring field theory in linear dilaton background,''
  arXiv:1605.04666 [hep-th].

\bibitem{Ishibashi:2016jno}
  N.~Ishibashi and K.~Murakami,
 ``Multiloop Amplitudes of Light-cone Gauge NSR String Field Theory in Noncritical Dimensions,''
  arXiv:1611.06340 [hep-th].

\end{thebibliography}
\end{document}